\documentclass[12pt]{article} 
\usepackage[sectionbib]{natbib}
\usepackage{array,epsfig,fancyheadings,rotating}
\usepackage{hyperref,xr} 
\usepackage{sectsty, secdot}
\sectionfont{\fontsize{12}{14pt plus.8pt minus .6pt}\selectfont}
\renewcommand{\theequation}{\thesection\arabic{equation}}
\subsectionfont{\fontsize{12}{14pt plus.8pt minus .6pt}\selectfont}

\usepackage{geometry}
\usepackage{amsmath}
\usepackage{amssymb}
\usepackage{amsfonts}
\usepackage{multirow}
\usepackage{amsthm}

\newtheorem{theorem}{Theorem}
\newtheorem{lemma}{Lemma}

\newtheorem{proposition}{Proposition}
\newtheorem{assumption}{Assumption}

\theoremstyle{definition}

\newtheorem{remark}{Remark}
\makeatletter 
\newcommand{\fdsy@scale}{1.0} 
\newcommand\fdsy@mweight@normal{Book} 
\newcommand\fdsy@mweight@small{Book} 
\newcommand\fdsy@bweight@normal{Medium} 
\newcommand\fdsy@bweight@small{Medium} 
\DeclareFontFamily{U}{FdSymbolC}{} 
\DeclareFontShape{U}{FdSymbolC}{m}{n}
{ <-7.1> s * [\fdsy@scale] FdSymbolC-\fdsy@mweight@small
	<7.1-> s * [\fdsy@scale] FdSymbolC-\fdsy@mweight@normal 
}{} 
\DeclareFontShape{U}{FdSymbolC}{b}{n}
{ <-7.1> s * [\fdsy@scale] FdSymbolC-\fdsy@bweight@small
	<7.1-> s * [\fdsy@scale] FdSymbolC-\fdsy@bweight@normal
}{} 
\DeclareSymbolFont{arrows}{U}{FdSymbolC}{m}{n}
\SetSymbolFont{arrows}{bold}{U}{FdSymbolC}{b}{n} 
\DeclareMathSymbol{\upvDash}{\mathrel}{arrows}{233} 
\let\Vbar\upvDash 
\DeclareMathSymbol{\upmodels}{\mathrel}{arrows}{237} 
\makeatother

\begin{document}


\renewcommand{\baselinestretch}{2}

\markright{ \hbox{\footnotesize\rm Statistica Sinica
}\hfill\\[-13pt]
\hbox{\footnotesize\rm
}\hfill }

\markboth{\hfill{\footnotesize\rm Yijiao Zhang and Zhongyi Zhu} \hfill}
{\hfill {\footnotesize\rm A DATA FUSION METHOD FOR QUANTILE TREATMENT EFFECTS} \hfill}

\renewcommand{\thefootnote}{}
$\ $\par


\fontsize{12}{14pt plus.8pt minus .6pt}\selectfont \vspace{0.8pc}
\centerline{\large\bf A DATA FUSION METHOD}
\vspace{2pt} 
\centerline{\large\bf FOR QUANTILE TREATMENT EFFECTS}
\vspace{.4cm} 
\centerline{Yijiao Zhang and Zhongyi Zhu } 
\vspace{.4cm} 
\centerline{\it Department of Statistics and Data Science, Fudan University}
 \vspace{.55cm} \fontsize{9}{11.5pt plus.8pt minus.6pt}\selectfont


\begin{quotation}
\noindent {\it Abstract:}
{\bf }
With the increasing availability of datasets, developing data fusion methods to leverage the strengths of different datasets to draw causal effects is of great practical importance to many scientific fields. In this paper, we consider estimating the quantile treatment effects using small validation data with fully-observed confounders and large auxiliary data with unmeasured confounders. We propose a Fused Quantile Treatment effects Estimator (FQTE) by integrating the information from two datasets based on doubly robust estimating functions. We allow for the misspecification of the models on the dataset with unmeasured confounders. Under mild conditions, we show that the proposed FQTE is asymptotically normal and more efficient than the initial QTE estimator using the validation data solely. By establishing the asymptotic linear forms of related estimators, convenient methods for covariance estimation are provided. Simulation studies demonstrate the empirical validity and improved efficiency of our fused estimators. We illustrate the proposed method with an application.

\vspace{9pt}
\par
\end{quotation}\par

\def\thefigure{\arabic{figure}}
\def\thetable{\arabic{table}}

\renewcommand{\theequation}{\thesection.\arabic{equation}}

\fontsize{12}{14pt plus.8pt minus .6pt}\selectfont

\par
\lhead[\footnotesize\thepage\fancyplain{}\leftmark]{\fancyplain{}\rightmark}\rhead[]{\thepage}

\section{Introduction}
\label{sec:intro}

The increasing availability of datasets from multiple sources holds enormous promise for evaluating causal effects. With various datasets at hand, data fusion technology has become more and more important in many medical and biological applications. How to systematically combine multiple datasets sources in an attempt to leverage the strengths of different types of data to improve the estimating efficiency of causal effects is gathering notice from researchers. For example, there are data sources with large sample size, such as electronic health records, claims databases, disease data registries, and census data.  However, uncontrolled design mechanisms and limited information on baseline covariates may lead to confounding bias,  presenting a major threat to causal inference. In practice, there are also small validation datasets that include all possible confounders and provide detailed information for each individual, especially in some randomized controlled trials (RCTs) in the medical field.  A classic example is a two-phase study \citep{wang2009causal}, where less expensive covariates are measured for all subjects in the first phase and the detailed information is collected in the second phase only for a validation subset drawn from the full sample. Unfortunately, the validation datasets often suffer from limited sample size due to the limitation of cost.  Therefore, evaluating causal effects based solely on the validation datasets lacks efficiency.  Consequently, we are seeking estimators of higher efficiency while pursuing unbiasedness as well by integrating information from both types of datasets. 

In the literature on causal inference, a great much of attention has been paid to the average treatment effects (ATE). In addition, quantiles are also useful measures for detecting causal effects. Firstly, when the outcomes are distributed with heavy tails, the medians are more efficient than the means. Secondly, quantiles are more appropriate measures when the distributions of outcomes are skewed. Thirdly, quantiles can also provide a more detailed view of heterogeneous causal effects at different points. In particular, researchers or policy-makers may be more interested in the distributional impacts on the dispersion of the outcome or the lower or higher tail of the distributions of potential outcomes beyond the average effects of treatment.  

In this article, we consider the data fusion problem of estimating the quantile treatment effects (QTE), defined as the difference between the quantiles of the marginal potential distributions of the treatment and control responses. We focus on the case where there are two types of data sources. One is a validation dataset that includes the measurements of all the confounders but suffers from small sample size and the other is an auxiliary dataset that enjoys large sample size but has unmeasured confounders.

In the case where there are no unmeasured confounders, several works have been done on identifying and estimating the conditional or unconditional QTE; for example, \citet{firpo2007efficient}, \citet{zhang2012causal}, \citet{donald2014estimation}, to name a few. \citep{firpo2007efficient} proposed an inverse probability weighting (IPW) estimator based on a nonparametric power series estimator of the propensity score and showed that under regular conditions their IPW estimator is root $n$ consistent and achieves the semiparametric efficiency bound. \citep{zhang2012causal} proposed an outcome regression (OR) estimator and a parametric inverse probability weighting (IPW) estimator based on a pre-specified outcome model and propensity score model respectively. They also proposed a doubly robust (DR) estimator which is consistent if either the outcome model or the propensity score model is correctly specified. In these papers, the unconfoundedness treatment assignment and strict overlap assumption are assumed. However, in observational studies, the unconfoundedness assumption may be violated due to unmeasured confounders, which is also called endogeneity of the treatment variable in the economics literature. To deal with this, instrumental variable (IV) approaches are developed for identifying the average or quantile treatment effects, see, for example, \citet{imbens1994identification}, \citet{wuthrich2019closed} for details. Nevertheless, valid instrumental variables are often difficult to find in practice. 

A burgeoning literature on data fusion has explored the possibility of harmonizing evidence from multiple data sources for estimating causal effects. Refer to \citet{colnet2020causal} for a detailed review. When the unconfoundedness is not assumed in the auxiliary big dataset, several methods are developed to deal with the confounding bias. One line is to construct shrinkage estimators by combining unbiased and biased estimators \citep{rosenman2022propensity,cheng2021adaptive}, which achieve lower MSE than the initial unbiased estimator based solely on the validation data. Another line is to specify a parametric model for the confounding bias \citep{kallus2018removing,yang2020improved}. However, the bias model is difficult to be correctly specified, especially in the case of QTE estimation. 

Empirical likelihood approaches are also commonly used for integrating information from multiple data sources \citep{chatterjee2016constrained,zhang2020generalized}. \citet{chatterjee2016constrained} consider a constrained maximum likelihood estimator using summary-level information from an external study, which is shown to be more efficient than the estimator based only on internal sample data. However, their method focuses on the regression parameters and can not be applied directly to the causal inference framework. Besides, the empirical likelihood approach always needs heavy computation when the data size gets large.

Considerably less work is available in this literature for QTE estimation under the data fusion framework. To the best of our knowledge, the only one is by \citet{li2021efficient}, which proposed a general semiparametric framework for efficient estimation under data fusion, including the estimation of QTE. However, their estimation is based on the canonical gradient, which may be sophisticated when there are unmeasured confounders.

Under our framework, though the two types of data may not be combined directly, they may share some common information. A natural idea is to connect the shared common information in multiple data sources through a calibration technique \citep[see e.g.][]{wu2001model,lin2014adjustment}. More related to our work, \citet{yang2020combining} used a calibration idea to improve the efficiency of the initial estimators of ATE by projecting them to the difference between two error-prone estimators based on the validation data and the auxiliary data respectively. A similar idea is also used in \citet{cai2021coda} for developing the optimal decision rule. The key insight is that the differences should be consistent estimators for zero. However, due to the essential properties of quantiles, a direct application of the difference-based method for QTE estimation involves the estimation of covariance matrices with rather complicated forms. Bootstrap methods for variance estimation could be time-consuming since the estimators themselves are constructed based on estimated covariance. 

In this article, we propose a fused quantile treatment effects estimator (FQTE) by integrating the information from both the validation data and the auxiliary data through estimating functions. We break down our contribution as follows:

On the methodological side, we show how to connect the two datasets through estimating functions and project the initial estimators on them to obtain our FQTE. We rationalize our idea from three different perspectives. Firstly, we can treat the biased estimators for quantiles as summary-level information from the big main data and use the estimation equation of the summary-level information as moment conditions on the validation dataset to make calibration to our initial estimators. Secondly, these estimating functions can be interpreted as linear combinations of doubly robust rank scores, which preserve the main information for quantiles robustly. Thirdly, as these estimating functions are consistent estimators for zero as well, we show our method as a generalization of \citet{yang2020combining} by reformulating the difference-based estimators for zero thereof. 

On the theoretical side, there are two core results. Firstly, we establish the asymptotic linear representations of the initial QTE estimators as well as those estimating functions we project on. This is the fundamental property that does not hold if we use the difference-based method in \citet{yang2020combining} for QTE estimation. Thanks to these asymptotic linear representations, convenient covariance estimation methods are provided to make our method easy to implement. To derive the asymptotic linear representations, the non-smooth estimating functions are dealt with via empirical process theory \citep[e.g.][Chap. 8]{kosorok2008introduction}. Secondly, consistency and asymptotic normality of our FQTE are established and we show that FQTEs enjoy efficiency gains. Besides, we also establish the consistency of our variance estimators based on the asymptotic variance.

By applying a missing mechanism to the datasets, we further extend our method to the cases where the validation sample may not be a random sample from the entire dataset, which is more reasonable in practice. Estimators and asymptotic results are provided based on unknown missing probabilities, which are assumed to be observed in \citet{yang2020combining}.

The rest of this paper is organized as follows. In Section \ref{sec:setup}, we give an exposition of the problem setup. We begin in Section \ref{sec:method} by proposing our data fusion method and then provide heuristic explanations. Section \ref{sec:theory} establishes the theoretical properties. The finite sample properties on simulated and real datasets are investigated in Section \ref{sec:simulation} and Section \ref{sec:application}. We conclude this paper with a discussion in Section \ref{sec:discussion}. 
\section{Setup and Basic Estimators}
\label{sec:setup}
\subsection{Basic Notations}
\label{subsec:notation}
We focus on the scenario where there is a validation dataset with fully-observed confounders and an auxiliary dataset with partially-observed confounders. We adopt the potential outcomes framework of Neyman and Rubin. See \citet{rubin1974estimating}. We focus on a binary treatment $T \in \{0,1\}$, which is an intervention of interest. Let $Y(t), t \in \{0,1\}$ denote the potential outcomes, which we interpret as the outcome had the individual assigned to treatment $t$. We assume the consistency assumption always holds, that is, the observed outcome $Y$ with an assigned treatment $T$ equal to $t$ equal to its potential outcome $Y(t)$, i.e., $Y=Y(T)=TY(1)+(1-T)Y(0)$. Let $X$ denote a $p_x$-dimensional vector of pre-treatment baseline covariates with support $\mathbb{X}$, $S$ a $p_s$-dimensional vector of pre-treatment baseline covariates with support $\mathbb{S}$ and $R$ a binary indicator for being in the validation sample or not ($R_i=1$ if the $i$th individual is in the validation data and $R_i=0$ otherwise). We model each individual in the observed data by a random tuple $(Y(1),Y(0),T,X^{\top},S^{\top},R)$ drawn from a superpopulation $\mathcal{P}$. We use $\mathrm{pr}$ to denote the distribution under $\mathcal{P}$ and denote $E$ as the expectation operator under $\mathrm{pr}$. The basic information $X$ is observed for all individuals, but the more detailed information $S$ is observed only on a subset of individuals. Denote the full-observed information as $O=(Y,T,X^{\top},S^{\top})$ and the partially-observed information as $U=(Y,T,X^{\top})$ with support $\mathbb{U}$. The validation dataset $\{O_i=(Y_i,T_i,X_i^{\top}, S_i^{\top}):i=1,...,n\}$ consist of $n$ identically and independently distributed (i.i.d.) observations , while the auxiliary data $\{U_i=(Y_i,T_i,X_i^{\top}):i=n+1,...,N\}$ consist of $m=N-n$ i.i.d observations without $S$. Define $\nu_n=n/N$ as the sample ratio between the validation data and the entire observed data and $\nu_n \rightarrow \nu \in [0,1)$ as $n \rightarrow \infty$. The entire observed data could thus be formulated as $\{D_i=(R_i,Y_i,T_i,X_i^{\top},R_iS_i^{\top}):i=1,...,N\}$. Define the index set $\mathcal{V}=\{1,...,n\}$ and $\mathcal{O}=\{1,...,N\}$. 

Define $F_t(y\mid X,S)=\mathrm{pr}(Y \leq y| T=t,X,S)$ and $F_t(y\mid X)=\mathrm{pr}(Y \leq y\mid T=t,X)$ as the conditional distribution of the observed outcome given the fully-observed covariates $(X,S)$ and partially-observed covariates $X$ respectively. Denote the conditional probability of the treatment as 
\begin{equation*}
e(X,S)=\mathrm{pr}(T=1\mid X,S),\quad e(X)=\mathrm{pr}(T=1\mid X).
\end{equation*}
The former is known as the propensity score in the causal inference literature. We call the latter one the pseudo propensity score as it does not include the information of the unmeasured confounders $S$.

Now we consider the quantile treatment effects. Denote $p$ as the quantile level and $F_t(\cdot)$ as the marginal cumulative distribution function of $Y(t)$. Formally, for any given quantile level $p \in (0,1)$, the $p$th quantile treatment effect is defined as
\begin{equation*}\label{qte}
\Delta_{p}=q_{1, p}-q_{0, p},
\end{equation*}
where $q_{t, p}=\inf \{q: F_t(q) \geq p\}$ is the $p$th quantile of $F_t(\cdot)$.

\subsection{Estimators Using the Fully-observed Validation Data}
\label{subsec:full}
The following are classical assumptions for identifying the quantile treatment effects.
\begin{assumption}[Ignorability]
	\label{assumption1}
	$Y(t) \upvDash T \mid (X,S)$ for $t=0,1.$ 
\end{assumption}

\begin{assumption}[Overlap]
	\label{assumption2}
	There exist constants $c_{1}$ and $c_{2}$ such that with probability $1,0<c_{1} \leq e(X, S) \leq c_{2}<1$.
\end{assumption}

The distribution of the potential outcomes can be identified under Assumptions \ref{assumption1} and \ref{assumption2} and classical estimators have been developed for estimating the QTE, including the outcome regression (OR) \citep{zhang2012causal}, inverse probability weighting (IPW), doubly robust (DR) estimators \citep{firpo2007efficient,zhang2012causal,donald2014estimation}. We assume that Assumptions \ref{assumption1} and \ref{assumption2} hold hereafter. Therefore, we can obtain an initial estimator using the validation data only.

Let $G_t(y\mid X,S;\theta_t)$ be a parametric working outcome regression (OR) model for $F_t(y\mid X,S)$, for example, a normal linear model after a Box-Cox transformation. Let $e(X,S;\alpha)$ be a parametric working propensity score (PS) model for $e(X,S)$. A common choice would be a logistic regression model. Let $\hat\theta^{\mathcal{V}}_{t}$ and $\hat\alpha^{\mathcal{V}}$ be consistent estimators for the corresponding true parameters $\theta_t^{*}$ and $\alpha^{*}$ based on the validation sample, for example, the maximum likelihood estimator (MLE). For simplicity, we omit the superscript $\mathcal{V}$ and denote the estimators as $\hat\theta_{t}$ and $\hat{\alpha}$ hereafter with no ambiguity. Define the weights $ w^{*}_{1, i}={T_{i}}/{e\left(X_{i},S_{i} ; {\alpha^{*}}\right)}$, $ w^{*}_{0, i}={(1-T_{i})}/{(1-e\left(X_{i},S_{i} ; {\alpha^{*}}\right))}$ and the estimated weights $ \hat w_{t, i}$ with ${\alpha}^{*}$ in $w^{*}_{t, i}$ replaced by its estimates $\hat{\alpha}$. Further denote $ {T}/{e\left(X,S ; {\alpha^{*}}\right)}$ as $w^{*}_{1}$, and ${(1-T)}/{(1-e(X,S ; {\alpha^{*}}))}$ as $w^{*}_{0}$.

\begin{assumption}[Outcome Model]
	\label{assumption3}
	The parametric model $G_t(y\mid X,S;\theta_t)$ is a correct specification for $F_t(y\mid X,S)$, for $t=0,1$, that is, $F_t(y\mid X,S)=G_t(y\mid X,S;\theta_t^{*})$, where $G$ is a known function and $\theta_{t}^{*}$ is the true model parameter, for $t=0,1$.
\end{assumption}
Similar assumptions about the correct specification of conditional distribution have been proposed in \cite{zhang2012causal} and \cite{han2019general} for quantile estimation with missing data. We may relax it to the correct specification of the conditional quantile, which is discussed in Section \ref{subsec:assump3}.
\begin{assumption}[Propensity score model]
	\label{assumption4}
	The parametric model $e(X, S ; \alpha)$ is a correct specification for $e(X, S)$; that is, $e(X, S)=$ $e\left(X, S ; \alpha^{*}\right)$, where $\alpha^{*}$ is the true model parameter.
\end{assumption}

By modelling both the conditional distribution $F_t(y\mid X,S)$ and the propensity score $e(X,S)$, we arrive at the so-called doubly robust (DR) estimator \citep{zhang2012causal}. For simplicity, let $\eta_{t}=(\theta_{t},\alpha)$ denote the nuisance parameter,  with $\eta^{*}_{t}=(\theta^{*}_{t},\alpha^{*})$ being its true value and $\hat\eta_{t}=(\hat\theta_{t},\hat\alpha)$ being its estimator. Under Assumptions \ref{assumption3} or \ref{assumption4}, $q_{t,p} $ $(t=0,1)$ can be identified by 
$E\left\{\Psi^{}_t(O;q_{t,p},{\eta}_t^{*})\right\}=0$, where
\begin{equation}\label{dr}
\Psi_t(O;q_{t,p},{\eta}_t^{*})=w^{*}_{t}\left\{I{(Y\leqslant q_{t,p})}-G_{t}\left(q_{t,p} \mid X, S ;{\theta}^{*}_{t}\right)\right\}+G_{t}\left(q_{t,p} \mid X, S; {\theta}^{*}_{t}\right)-p.
\end{equation}
A DR estimator for $\Delta_{p}$ based on the validation sample is defined as $\hat{\Delta}^{\operatorname{\mathcal{V}}}_{p}$ $=\hat{q}^{\mathcal{V}}_{1,p}$ $-\hat{q}^{\mathcal{V}}_{0,p}$, where $\hat{q}^{\mathcal{V}}_{t,p}$ is an DR quantile estimator for $q_{t,p} (t=0,1)$, which is the solution to
\begin{equation}\label{emdr}
	{1}/{n} \sum_{i=1}^{n} \Psi_t(O_i;q,\hat{\eta}_{t})= 0,
\end{equation}
where ${\Psi_t(O_i;q,\hat{\eta}_{t})=\hat w_{t,i}\{I{(Y_{i} \leqslant q)}-G_{t}(q \mid X_{i}, S_{i} ;\hat{\theta}_{t})\}+G_{t}(q \mid X_{i}, S_{i}; \hat{\theta}_{t})-p.}$ 
	
The sum of weights $\sum_{i=1}^{n}\hat{w}_{t,i}$ converges to 1 as $n \rightarrow \infty$ but is generally different from 1 for any finite $n.$ For improved finite-sample performance, normalized weights can be calculated. The DR estimator is consistent if either the OR or the PS model is correctly specified. Moreover, the DR estimator is locally efficient when both outcome and propensity score models are correctly specified \citep{diaz2017efficient}. 

Under regular conditions, the DR estimator is also asymptotically linear in the sense of \citep{tsiatis_semiparametric_2007}. To be detailed, according to Theorem \ref{theorem1} in Section \ref{subsec:ALR}, we have for the DR quantile estimators that
\begin{equation}\label{linearqte}
n^{1/2}(\hat{q}^{\mathcal{V}}_{t,p}-q_{t,p})={1}/{n^{1/2}}\sum_{i=1}^{n}\psi_t(O_i;q_{t,p},\eta_{t}^{*})+o_p(1).
\end{equation}
where $\psi_t(O_i;q_{t,p},\eta_{t}^{*})$ is given in (\ref{INFL}) in Section \ref{subsec:ALR} and it is called the influence function of $\hat{q}^{\mathcal{V}}_{t,p}$. The asymptotic linear representations  provide us with great convenience for variance estimation. 

\subsection{Estimators Using the Partially-observed Entire Data}
\label{subsec:partial}
The initial estimators only use information from the validation sample which has a small sample size and hence lacks efficiency. That's why we need the auxiliary datasets with a large sample size to help improve the efficiency. However, the auxiliary dataset does not include detailed information about $S$ and hence may lead to confounding bias. Nevertheless, we may treat $X$ as all the confounders and use the entire data with only $U=(Y, T, X^{\top})$ to obtain quantile and QTE estimators, following the same estimating procedure as we do on the validation dataset.

To be specific, we may consider using the same working models as that in Section \ref{subsec:full} for the conditional distribution $F_t(y\mid X)$ (for example, both normal linear) and the pseudo propensity score $e(X)$ (for example, both logistic), which may be misspecified. Denote them as $\tilde{G}(X;\theta_t^{\operatorname{Conf}})$ and $\tilde{e}(X;\alpha^{\operatorname{Conf}})$ respectively. Here ``$\mathrm{Conf}$'' is short for ``confounded''. Let $\hat\theta_t^{\operatorname{Conf},\mathcal{O}}$ and $\hat\alpha^{\operatorname{Conf},\mathcal{O}}$ be the corresponding MLEs based on the entire sample, with probability limits $\theta_t^{\operatorname{Conf},*}$ and $\alpha^{\operatorname{Conf},*}$ \citep{white1982maximum}. For simplicity, we omit the superscript "$\mathcal{O}$" and denote the estimators as $\hat\theta^{\operatorname{Conf}}_{t}$ and $\hat\alpha^{\operatorname{Conf}}$ hereafter. Define the weights $ z^{*}_{1, i}={T_{i}}/{e\left(X_{i} ; {\alpha^{\operatorname{Conf},*}}\right)}$, $ z^{*}_{0, i}={(1-T_{i})}/{(1-e\left(X_{i} ; {\alpha^{\operatorname{Conf},*}}\right))}$ and the estimated weights $ \hat {z}_{t, i}$ with ${\alpha^{\operatorname{Conf},*}}$ in $z^{*}_{t, i}$ replaced by its estimates $\hat\alpha^{\operatorname{Conf}}$. Further denote ${T}/{e\left(X ; {\alpha^{\operatorname{Conf},*}}\right)}$ as $ z^{*}_{1}$ and ${(1-T)}/{(1-e(X ; {\alpha^{\operatorname{Conf},*}}))}$ as $z^{*}_{0}$. Similarly, let $\eta_{t}^{\operatorname{Conf}}=(\theta_{t}^{\operatorname{Conf}},\alpha^{\operatorname{Conf}})$ denote the nuisance parameter, with $\eta_{t}^{\operatorname{Conf},*}=(\theta_{t}^{\operatorname{Conf},*},\alpha^{\operatorname{Conf},*})$ being its probability limit and $\hat\eta_{t}^{\operatorname{Conf}}=(\hat\theta_{t}^{\operatorname{Conf}},\hat\alpha^{\operatorname{Conf}})$ being its consistent estimator.

We may construct empirical estimation equations similar to (\ref{emdr}) to obtain DR quantile estimators $\hat q_{t,p}^{\operatorname{Conf}}$ subject to unmeasured confounding, by solving the equation (in $q$) 
\begin{equation}\label{Confdr}
{1}/{N}\sum_{i=1}^{N}\phi_t(U_i;q,{\hat\eta}^{\operatorname{Conf}}_{t})=0,
\end{equation}
where ${\phi_t(U_i;q,{\hat\eta}^{\operatorname{Conf}}_{t})=\hat{z}_{t,i}\{I{(Y_{i} \leqslant q)}-\tilde G_{t}(q \mid X_{i} ;\hat{\theta}^{\operatorname{Conf}}_{t})\}+\tilde G_{t}(q \mid X_{i}; \hat{\theta}^{\operatorname{Conf}}_{t})-p.}$ 

In fact, the parameter of which $\hat q_{t,p}^{\operatorname{Conf}}$ estimates , denoted as $q_{t,p}^{\operatorname{Conf}}$, can be identified by the following equation on the partially-observed data
\begin{equation}\label{Confemdr}
E\left\{\phi_t(U;q_{t,p}^{\operatorname{Conf}},\eta_t^{\operatorname{Conf},*})\right\}=0.
\end{equation}

We call $q_{t,p}^{\operatorname{Conf}}$ as the pseudo quantile. To explain it, if the unconfoundedness assumption also holds with the partially-observed confounders $X$, we have that $q_{t,p}^{\operatorname{Conf}}$ is equal to $q_{t,p}$ if either $\tilde{G}(X;\theta_t^{\operatorname{Conf}})$ is correctly specified for $F_t(y\mid X)$ or $\tilde{e}(X;\alpha^{\operatorname{Conf}})$ is correctly specified for $e(X)$. Then we will obtain $N^{1/2}$-consistent estimators for the quantiles and then the QTE, which are more efficient than the initial estimators. However, due to the unmeasured confounders $S$, $q_{t,p}^{\operatorname{Conf}}$ can be different from $q_{t,p}$, which leads $\hat q_{t,p}^{\operatorname{Conf}}$ to be biased estimators for our interested parameter $q_{t,p}$. Consequently, the QTE estimators using $X$ only are often inconsistent. 

The question is how to integrate the unbiased but inefficient estimator $\hat{q}^{\mathcal{V}}_{t,p}$ in Section \ref{subsec:full} and the more efficient but biased estimator $\hat q_{t,p}^{\operatorname{Conf}}$ here to improve the efficiency of the initial estimators while pursuing consistency as well. This is what we do in the next section.
\section{Method}
\label{sec:method}
\subsection{Proposed Method}
\label{subsec:method}
Now we consider using the calibration technique to fuse the two datasets to draw inference for the QTE. The key idea is to connect the two datasets through estimating functions and then take projection. 

Though the two types of data may not be fused directly, they may share some common information. A natural idea is to connect the shared common information in multiple data sources through a calibration technique \citep[see e.g.][]{lin2014adjustment}.
To make a calibration, we need assumptions about the shared information on the two datasets. The following assumption is classical in the missing data literature.
\begin{assumption}[Missing Completely at Random, MCAR]\label{assumption5}
	$R \Vbar (Y,T,X,S)$.
\end{assumption} 
We further extend this MCAR assumption to a weaker missing at random (MAR) assumption ($R \upvDash S | (Y,T,X)$) in the supplementary materials Section S2, which allows the selection of the validation sample to depend on a probability design. Under Assumption \ref{assumption5}, $\{(Y_i,T_i,X_i^{\top}),i=1,...,N\}$ are i.i.d. samples. Then we have that the equation $E\left\{\phi_t(U;q^{\operatorname{Conf}}_{t,p},\eta^{\operatorname{Conf}}_{t})\right\}=0$ will also hold on the validation sample, which motivates us to connect the two datasets through the estimating functions
\begin{equation}\label{Ct}
\hat{C}_t={1}/{n}\sum_{i=1}^{n}{\phi_t(U_i;\hat{q}_{t,p}^{\operatorname{Conf}},\hat{\eta}^{\operatorname{Conf}}_t)},
\end{equation}
where $\phi_t(U_i;\cdot,\cdot)$ is defined in (\ref{Confdr}) in Section \ref{subsec:partial}. Note that (\ref{Ct}) integrates the information from the auxiliary data through ${\hat q}^{\operatorname{Conf}}_{t,p}$ and ${\hat\eta}^{\operatorname{Conf}}_{t}$ as well as the information from the validation data through $\{U_i\}_{i=1}^{n}$.

Similar to (\ref{linearqte}), under regular conditions given in detail in Section \ref{subsec:ALR}, we can also establish the asymptotical linear representations for $\hat C_t$ as
\begin{equation}\label{linearCt}
\hat{C}_t= {1}/{n}\sum_{i=1}^{n} \phi_t(U_i;q_{t,p}^{\operatorname{Conf},*},\eta^{\operatorname{Conf},*}_{t})-{1}/{N}\sum_{i=1}^{N} \phi_t(U_i;q_{t,p}^{\operatorname{Conf},*},\eta^{\operatorname{Conf},*}_{t})+o_p({n^{-1/2}}),
\end{equation}
which also implies that $\hat{C}_t$ is a consistent estimator for zero. Here we don't need any assumptions on the correct specification of models on the joint distribution of $U=(A,X,Y)$. We allow both the working models $F(y\mid X)$ and $e(X)$ to be misspecified. For similicity, denote  $\phi_{t}(U_i;q^{\operatorname{Conf},*}_{t,p},\eta_{t}^{\operatorname{Conf},*})$ as $\phi_{t,i}$ and $\psi_t(O_i;q_{t,p},\eta^{*}_{t})$ as $\psi_{t,i}$ for $t=0,1$. Combining (\ref{linearqte}) and (\ref{linearCt}), the next proposition models the asymptotic joint distribution of $\hat \Delta^{\mathcal{V}}_{p}$ and $\hat{C}_t$.

\begin{proposition}\label{proposition1}
	Under the assumptions in Theorem \ref{theorem1} and Lemma \ref{lemma1} in Section \ref{subsec:ALR}, as $n \rightarrow \infty$, then
	\begin{equation}\label{n1qte}
	n^{1 / 2}\left(\begin{array}{c}
	\hat{\Delta}^{\mathcal{V}}_{p}-\Delta_{p} \\  \hat{C}
	\end{array}\right) {\longrightarrow} \mathcal{N}\left\{0,\left(\begin{array}{cccc}
	\sigma_{\mathcal{V}}^{2} & {\varrho}^{\top} \\
	\varrho&\Sigma_{\mathrm{ep}}
	\end{array}\right)\right\}, 
	\end{equation}
	in distribution, where $\hat{C}=(\hat{C}_1^{\top},\hat{C}_0^{\top})^{\top}$, $\sigma_{\mathcal{V}}^{2}=\operatorname{var}(\psi_{1,i}-\psi_{0,i})$, $\varrho=(1-\nu) \operatorname{cov}(\psi_{1,i}-\psi_{0,i}, (\phi_{1,i}^{\top},\phi_{0,i}^{\top})^{\top})$,  $\Sigma_{\mathrm{ep}}=\left(\begin{array}{cccc}
	\Sigma_{1} & \Sigma_{01}^{\top} \\
	\Sigma_{01}&\Sigma_{0}
	\end{array}\right)$ with $\Sigma_{01}=(1-\nu)\operatorname{cov}(\phi_{0,i},\phi_{1,i})$ and $\Sigma_{t}=(1-\nu)$$\operatorname{var}(\phi_{t,i})$, for $t=0, 1$.
\end{proposition}

\begin{remark}
	\label{remark1}
	Proposition \ref{proposition1} is analogous to Theorem 1 in \citet{yang2020combining} for estimating ATE, which is the fundamental part of our method.
\end{remark}

By projecting $\hat{\Delta}^{\mathcal{V}}_{p}$ to $\hat{C}$, we can obtain our fused QTE estimator (FQTE)
\begin{equation}\label{estqte}
\hat{\Delta}_{p}=\hat{\Delta}^{\mathcal{V}}_{p}-\hat{\varrho}^{\top} \hat{\Sigma}_{\mathrm{ep}}^{-1}\hat{C},
\end{equation}
where $\hat{\varrho}$ and  $\hat{\Sigma}_{\mathrm{ep}}$ are corresponding consistent estimators for $\varrho$ and $\Sigma_{\mathrm{ep}}$. The construction of consistent covariance estimators $(\hat{\varrho},\hat{\Sigma}_{\mathrm{ep}})$ in (\ref{estqte}) will be discussed in the Section \ref{subsec:var}. We assume that $\Sigma_{\mathrm{ep}}$ is positive definite, which is similarly assumed in \cite{yang2020combining}.

\begin{remark} 
	\label{remark2}
	We can also integrate the information of pseudo quantiles at different orders to draw inference about the QTE at the $p$th order. Consider a $d$-dimensional vector $q^{\mathrm{Conf}}_t$, $t=0, 1$, with the $k$th dimension be the $p_k$th ($k=1,\ldots,d$) order pseudo quantile $q^{\mathrm{Conf}}_{t,p_k}$ identified by the equations $E\phi_{t,k}(U_i;q^{\mathrm{Conf}}_{t,p_k},{\eta}^{\operatorname{Conf},*}_{t})=0, t=0,1,$ where $\phi_{t,k}(U_i;q^{\mathrm{Conf}}_{t,p_k},{\eta}^{\operatorname{Conf},*}_{t})$ equals
	$$
	{z}^{*}_{t,i}\left\{I{(Y_{i} \leqslant q^{\mathrm{Conf}}_{t,p_k})}-\tilde G_{t}(q^{\mathrm{Conf}}_{t,p_k} \mid X_{i} ;{\theta}^{\operatorname{Conf},*}_{t})\right\}+\tilde G_{t}(q^{\mathrm{Conf}}_{t,p_k} \mid X_{i}; {\theta}^{\operatorname{Conf},*}_{t})-p_k
	$$
	Then we will obtain two $d$-dimensional vector $\hat{C}_1$ and $\hat{C}_0$. We denote hereafter the pseudo quantiles $(p_1,p_2,\ldots,p_d)$ chosen for calibration as $p_{\mathrm{cal}}$. 
\end{remark}

\subsection{Heuristic explanation}
\label{subsec:exp}
The construction of $\hat{C}_t$ is motivated by the usage of summary-level information in the data integration literature as well as the role of rank scores in the quantile regression analysis.

\citet{zhang2020generalized} also proposed an empirical likelihood approach for data integration, using the summary-level data to make constraints on moments on the validation sample. However, they focus on the regression analyses and the empirical likelihood approach cannot be directly applied here in the causal inference framework. Here we treat $\hat{q}_{t,p}^{\operatorname{Conf}}$ and $\hat{\eta}^{\operatorname{Conf}}_t$ as summary-level data obtained from the entire sample without the detailed confounders $S$. Equation (\ref{Ct}) is just obtained from the moment conditions based on the summary-level information. When the entire sample size $N$ is extremely large, the uncertainty in $\hat{q}_{t,p}^{\mathrm{Conf}}$ and $\hat{\eta}^{\mathrm{Conf}}_t$ can be ignored, therefore, we can simply treat them as the true parameters ${q}_{t,p}^{\mathrm{Conf},*}$ and ${\eta}^{*}_t$. The right side of (\ref{Ct}) is simply replacing the expectation in the left side of (\ref{Confemdr}) with the empirical measure based on the validation sample. 

We can also interpret $\hat{C}_t$ as doubly robust rank scores. Substituting $\phi_{t}$ in (\ref{Ct}) with its complete expression in (\ref{Confdr}), we obtain that 
\begin{equation}\label{rs}
\hat{C}_t={1}/{n}\sum_{i=1}^{n}\left[\hat{z}_{t,i}\left\{I{(Y_{i} \leqslant \hat q^{\operatorname{Conf}}_{t,p})}-\tilde G_{t}(\hat q^{\operatorname{Conf}}_{t,p} \mid X_{i} ;\hat{\theta}^{\operatorname{Conf}}_{t})\right\}+\tilde G_{t}(\hat q^{\operatorname{Conf}}_{t,p} \mid X_{i}; \hat{\theta}^{\operatorname{Conf}}_{t})-p\right]
\end{equation}
Here $I{(Y_{i} \leqslant \hat q^{\operatorname{Conf}}_{t,p})}-p$ serves as the rank score in the quantile regression literature \citep{koenker_2005} up to a constant involving the marginal density of $Y_i$. Consequently, it preserves the main information contained in $q^{\operatorname{Conf}}_{t,p}$. Here in (\ref{rs}), $\hat{C}_t$ can be interpreted as a linear combination of doubly robust rank scores where a pseudo propensity score model and an outcome model are included to improve its robustness. \citet{giessing2021Debiased} use the information in rank scores to debias the conditional quantile treatment effects, while here we use it to integrate information from two datasets.

Our method is also closely related to that in \citet{yang2020combining}. They consider the same data configuration as our paper but focus on estimating the ATE, $\tau=E\{Y(1)-Y(0)\}$. The common ground between our method and theirs is that we both project the initial estimators on a consistent estimator for zero. However, a significant difference is that we connect the two datasets through estimating functions rather than simple differences to produce a consistent estimator for zero, which makes it easily adapted to the estimation of QTE. Specifically, suppose that $\tau$ is identified by $E\left\{\varphi(U;\tau,\gamma)\right\}=0$ with a nuisance $\gamma$. \citet{yang2020combining} proposed to project the initial estimator on the difference between two error-prone ATE estimators,  $\hat{\tau}^{\mathcal{V}}_{\mathrm{ep}}$ and $\hat{\tau}^{\mathcal{O}}_{\mathrm{ep}}$, which are obtained by solving the empirical version of $E\left\{\varphi(U;\tau,\gamma)\right\}=0$ based on the validation data and the entire data separately. Extended to our QTE case, it is equivalent to project on the difference $\hat{C}_{\mathrm{ep}}$ between $\hat{\Delta}^{\mathcal{V},\mathrm{Conf}}_{p}=\hat{q}^{\mathcal{V},\mathrm{Conf}}_{1,p}-\hat{q}^{\mathcal{V},\mathrm{Conf}}_{0,p}$ and $\hat{\Delta}^{\mathcal{O},\mathrm{Conf}}_{p}=\hat{q}^{\mathcal{O},\mathrm{Conf}}_{1,p}-\hat{q}^{\mathcal{O},\mathrm{Conf}}_{0,p}$, obtained by solving the empirical version of (\ref{Confemdr}) based on two samples separately. Suppose the estimators satisfy
\begin{equation}\label{n1qtediff}
n^{1 / 2}\left(\begin{array}{c}
\hat{\Delta}^{\mathcal{V}}_{p}-\Delta_{p} \\  \hat{C}_{\mathrm{ep}}
\end{array}\right) {\longrightarrow} \mathcal{N}\left\{0,\left(\begin{array}{cccc}
\sigma_{\mathcal{V}}^{2} & {\Gamma}^{\top} \\
\Gamma&V
\end{array}\right)\right\}. 
\end{equation}
We can then obtain a difference-based estimator $\hat{\Delta}^{\mathrm{diff}}_{p}=\hat{\Delta}^{\mathcal{V}}_{p}-\hat{\Gamma}^{\top} \hat{V}^{-1}\hat{C}_{\mathrm{ep}}$ given consistent estimators $\hat{\Gamma}$ and $\hat{V}$.

Unfortunately, due to the essential properties of quantiles, the covariance matrices $\Gamma$ and $V$ are rather complicated.  Heuristically, due to the unmeasured $S$ in the partially-observed data, there are unknown terms related to $S$ appeared simultaneously in $\Gamma$ and $V$: the true propensity score $\mathrm{pr}(T=1|X,S)$, the conditional distribution $F_t(y|X,S)$, as well as the conditional density $f_t(y|X,S)$. Estimation of these terms can be complicated and unstable without additional model assumptions, especially when the validation sample size is relatively small. Although bootstrap method can be used to estimate $\Gamma$ and $V$, it can be time-consuming, especially when the entire data size is large, which is exactly the scenario we consider here. See the supplementary materials Section S3.3 for more details and discussions.

However, inspired by the perspective of summary-level data, we can reformulate $\hat{\tau}_{\mathrm{ep}}=\hat{\tau}^{\mathcal{V}}_{\mathrm{ep}}-\hat{\tau}^{\mathcal{O}}_{\mathrm{ep}}$ in \citet{yang2020combining} as $\hat{\tau}_{\mathrm{ep}}={1}/{n}\sum_{i=1}^{n}\varphi(U_i;\hat{\tau}^{\mathcal{O}}_{\mathrm{ep}},\hat\gamma^{\mathcal{O}})$ by treating $\hat{\tau}^{\mathcal{O}}_{\mathrm{ep}}$ and $\hat\gamma^{\mathcal{O}}$ as summary-level data from the dataset with unmeasure confounders, where $\hat\gamma^{\mathcal{O}}$ is an estimator for $\gamma$ using the entire data. Thus, the estimating-function-based connection can be regarded as a generalization of the difference-based connection in \citet{yang2020combining}, which, in the case of QTE estimation, leads to $\hat{C}_t$ as in (\ref{Ct}) with an asymptotical linear representation in (\ref{linearCt}). Combined with the linear form of the initial estimators in (\ref{linearqte}), the covariance matrix can be easily estimated. As we can see from the simulation results in the supplementary materials Section S4.3, besides less computation time, our proposed FQTEs also enjoy better performance than their difference-based estimators.
\section{Theoretical Guarantees}
\label{sec:theory}
\subsection{Asymptotic Linear Representation}
\label{subsec:ALR}
For the theoretical analysis, let us introduce some additional notations. We use $\xi^{\otimes 2}=\xi \xi^{\top}$ for a vector or matrix $\xi$. For the outcome model, let $L_{t}\left(Y, T, X, S ; \theta_{t}\right)$ be the estimating function for $\theta_{t}$, for $t=0,1$. For the propensity score model, let $h(T, X, S ; \alpha)$ be the estimating function for $\alpha$. Moreover, let $\Sigma_{\alpha} =E\left\{h^{\otimes 2}(T, X, S ; \alpha)\right\}$ be the Fisher information matrix for $\alpha$ in the propensity score model.  

Simply denote $e_{i}^{*}=e\left(X_{i}, S_{i};\alpha^{*}\right), \dot{e}_{i}^{*}=\partial e\left(X_{i}, S_{i};\alpha^{*}\right) / \partial \alpha^{\mathrm{T}}$, $h_{i}^{*}=
h\left(T_{i},\right.$ $ X_{i}, S_{i}; \alpha^{*}\left.\right)$, $\dot{h}_{i}^{*}=$ $\partial h\left(T_{i}, X_{i}, S_{i} ;\alpha^{*}\right)/$ $\partial {\alpha}^{\mathrm{T}} $, $G_{t i}^{*}=G_{t}\left(q_{t,p}\mid X_{i}, S_{i} ; \theta_{t}^{*}\right), \dot{G}_{t i}^{*}=\partial G_{t}\left(q_{t,p}\mid X_{i}, S_{i} ; \theta_{t}^{*}\right)/\partial \theta_{t}^{\mathrm{T}}$, $L_{t i}^{*}=L_{t}\left(Y_{i} ,T_{i}, X_{i}, S_{i}; \theta_{t}^{*}\right)$, and $\dot{L}_{t i}^{*}=\partial L_{t}\left(Y_{i} ,T_{i},\right.$ $ X_{i}, S_{i};\theta_{t}^{*}\left.\right) / \partial \theta_{t}^{\mathrm{T}}$ for $t=0,1$. Besides, we denote the density of $Y(t)$ as $f_t(y)$, for $t=0,1$. Now we establish the consistency and asymptotic normality of the DR QTE estimators. 
\begin{theorem}\label{theorem1}
	Under Assumptions \ref{assumption3} or \ref{assumption4}, \ref{assumption5}, and Condition S3.1 in the supplementary materials Section S3.1, the DR quantile estimators $\hat q^{\mathcal{V}}_{1,p}$ and $\hat q^{\mathcal{V}}_{0,p}$ obtained by solving (\ref{emdr}) are asymptotically normal, and (\ref{linearqte}) holds with the influence function 
	\begin{equation}\label{INFL}
	\begin{aligned}
	\psi_{1}(O_i;q_{1,p},\eta^{*}_{1})&=-{1}/{f_1(q_{1,p})}\left[{T_i\left\{I(Y_i\leq q_{1,p})-G_{1i,p}^{*}\right\}}/{e_i^{*}}+G_{1i,p}^{*}-p \right.\\&\left.-E\left\{\left(1-{T_i}/{e_i^{*}}\right)\dot{G}_{1i,p}^{*}\right\} (E\dot{L}_{1i}^{*})^{-1} L_{1i}^{*}-H_1\Sigma_{\alpha}^{-1}h_i^{*}\right],\\\psi_{0}(O_i;q_{0,p},\eta^{*}_{0})&=-{1}/{f_0(q_{0,p})}\left[{(1-T_i)\left\{I(Y_i\leq q_{t,p})-G_{0i,p}^{*}\right\}}/{(1-e_i^{*})}+G_{0i,p}^{*}-p \right.\\&\left.-E\left\{\left(1-{(1-T_i)}/{(1-e_i^{*})}\right)\dot{G}_{0i,p}^{*}\right\} (E\dot{L}_{0i}^{*})^{-1} L_{0i}^{*}-H_0\Sigma_{\alpha}^{-1}h_i^{*}\right],
	\end{aligned}
	\end{equation}
	respectively, where 
	\begin{equation*}
	\begin{aligned}
	H_1&=E\left[ {T_i\left\{I(Y_i\leq q_{1,p})-G_{1i,p}^{*}\right\}\dot{e}_i^{*}}/{{e_i^{*}}^2} \right],\\
	H_0&=-E\left[ {(1-T_i)\left\{I(Y_i\leq q_{0,p})-G_{0i,p}^{*}\right\}\dot{e}_i^{*}}/{{(1-e_i^{*})}^2} \right].
	\end{aligned}
	\end{equation*}
	Consequently, the DR QTE estimator $\hat{\Delta}_{p}$ is asymptotic linear with influence function $\psi_{1}(O_i;q_{1,p},\eta^{*}_{1})-\psi_{0}(O_i;q_{1,p},\eta^{*}_{1})$.
\end{theorem}
Similarly, we establish the asymptotic linear representations of $\hat{C}_t$. 
\begin{lemma}\label{lemma1}
	Under Assumption \ref{assumption5} and Condition S3.2 in the supplementary materials Section S3.1, we have
	\begin{equation*}
	\hat{C}_t= {1}/{n}\sum_{i=1}^{n} \phi_t(U_i;q_{t,p}^{\operatorname{Conf},*},\eta^{\operatorname{Conf},*}_{t})-{1}/{N}\sum_{i=1}^{N} \phi_t(U_i;q_{t,p}^{\operatorname{Conf},*},\eta^{\operatorname{Conf},*}_{t})+o_p({1}/{n^{1/2}}),
	\end{equation*}
	for $t=0,1$. That is, the asymptotic linear representation in (\ref{linearCt}) holds.
\end{lemma}

\begin{remark}
	\label{remark3}
	As we mentioned before, here we allow both $\tilde{G}(X;\theta_t^{\operatorname{Conf}})$ and $\tilde{e}(X;\alpha^{\operatorname{Conf}})$ to be misspecified for $F_t(y\mid X)$ and $e(X)$ respectively, this is analogous to that in \citep{yang2020combining}.
\end{remark}

\subsection{Efficiency Gains}
\label{subsec:gain}
The following theorem shows that our FQTE can improve the efficiency of the initial QTE estimators. 
\begin{theorem}[Asymptotic Normality]\label{theorem2}
	Under the assumptions in Theorem \ref{theorem1} and Lemma \ref{lemma1}, given consistent estimators $(\hat{\varrho},\hat{\Sigma}_{\mathrm{ep}})$ for $({\varrho},{\Sigma}_{\mathrm{ep}})$, $\hat \Delta_{p}$ is consistent and
	\begin{equation*}
	n^{1 / 2}(\hat\Delta_{p}-\Delta_p) {\longrightarrow} \mathcal{N}\left(0, \sigma^2\right)
	\end{equation*}
	in distribution as $n \rightarrow \infty$, where ${\sigma}^2=\sigma_{\mathcal{V}}^2-{\varrho}^{\top} \Sigma_{\mathrm{ep}}^{-1} \varrho$. 
\end{theorem}

Note that $\Sigma_{\mathrm{ep}}$ is positive definite, which means the FQTE have efficiency gains of ${\varrho}^{\top} \Sigma_{\mathrm{ep}}^{-1} \varrho$ compared to the initial estimator, given nonzero $\varrho$. Theorem 2 also implies that the efficiency gains increase with the covariance between the estimators for zero and the initial QTE estimators based solely on the validation dataset. The higher the covariance $\varrho$ is, the more efficiency gains we will obtain.

\subsection{Variance Estimation}
\label{subsec:var}
Now we discuss how to obtain consistent estimators for the asymptotic covariance matrix in (\ref{n1qte}). The asymptotic linear representations provide us this a convenient way to construct covariance estimators. Consistent estimators for $f_t(q_{t,p})$ are provided in the supplementary materials Section S1. Replacing the unknown terms $(\eta_{t},q_{t,p},f_t(\cdot))$ in $\phi_{t,i}$ and $(\eta_{t}^{\operatorname{conf}},q_{t,p}^{\operatorname{conf}})$ in $\psi_{t,i}$ by their correponding estimators $(\hat\eta_{t},\hat{q}_{t,p}^{\mathcal{V}},\hat{f}_t(\cdot))$ and $(\hat{\eta}_{t}^{\operatorname{Conf}},\hat{q}_{t,p}^{\operatorname{Conf}})$, using empirical measure in place of $E(\cdot)$, we can obtain estimators of $\phi_{t,i}$ and $\psi_{t,i}$ respectively, denoted by $\hat{\psi}_{t,i}$ and $\hat{\phi}_{t,i}$. Based on Proposition 1, we can estimate $(\varrho,\Sigma_{\mathrm{ep}},\sigma_{\mathcal{V}}^{2})$ by 
\begin{equation}\label{var1}
\begin{aligned}
\hat{\varrho}=\left(1-\nu_n\right) {1}/{n} \sum_{i=1}^{n} (\hat{\psi}_{1,i}-\hat{\psi}_{0,i}) (\hat{\phi}_{1,i}^{\top}, \hat{\phi}_{0,i}^{\top})^{\top}&,\quad \hat{\Sigma}_{t}=\left(1-\nu_n\right) {1}/{N} \sum_{i=1}^{N}\hat{\phi}_{t,i}\hat{\phi}_{t,i}^{\top}\\
\hat{\Sigma}_{01}=\left(1-\nu_n\right) {1}/{N} \sum_{i=1}^{N}\hat{\phi}_{0,i}\hat{\phi}_{1,i}^{\top}, 
&,\quad \hat\sigma_{\mathcal{V}}^{2}={1}/{n} \sum_{i=1}^{n}\left(\hat{\psi}_{1,i}-\hat{\psi}_{0,i}\right)^{2}.\\
\end{aligned}
\end{equation}
Based on the consistent estimators for the covariance matrix, we can obtain corresponding variance estimators for our calibrated estimators as
\begin{equation}\label{var2}
\hat{\sigma}^2=\hat\sigma_{\mathcal{V}}^{2}-\hat \varrho^{\top}\hat\Sigma_{\mathrm{ep}}^{-1} \hat \varrho. 
\end{equation}

Given a consistent density estimator, we can establish the consistency of the covariance and variance estimators. 
\begin{theorem}[Consistent Variance Estimators]\label{theorem3}
	Under Assumptions \ref{assumption3} or \ref{assumption4}, and regular conditions in the supplementary materials Section S3.1, given a consistent estimator $\hat{f}_t(y)$ for $t=0,1$, the covariance estimators $ (\hat \varrho, \hat{\Sigma}_{\mathrm{ep}}, \hat\sigma_{\mathcal{V}}^{2})$ in (\ref{var1}) are consistent for $(\varrho,\Sigma_{\mathrm{ep}},\sigma_{\mathcal{V}}^{2})$. Consequently, the variance estimator $\hat\sigma^{2}$ in (\ref{var2}) are consistent for $\sigma^{2}$.
\end{theorem}

\begin{remark}
	\label{remark4}
	All the results in Section \ref{sec:method} and \ref{sec:theory} are based on the MCAR assumption (Assumption \ref{assumption5}). Under a relaxed MAR assumption, where the validation data is no longer a random sample from the entire data, slight modifications to the construction of our FQTE are needed. Analogous results including asymptotic linear representations as well as efficiency gains under the MAR assumption are provided in the supplementary materials Section S2. Technique proofs of all the theorems and lemmas above are provided in the supplementary materials Section S3. 
\end{remark}

\section{Simulation}
\label{sec:simulation}
In this section, we	evaluate the empirical performance of our proposed FQTE under the MCAR assumption. Simulation results based on the relaxed MAR assumption as well as a comparison between our FQTE and the direct extension of the difference-based calibration method in \citep{yang2020combining} are provided in the supplementary materials Section S4. All the pre-treatment covariates are standardized to have mean 1 and variance 1. We consider a case with 4 confounders, only one commonly observed for the entire data and the rest three observed only for a subset of units. We first generate $W_{ki}$ from the uniform distribution $ \operatorname{Unif}(1-\sqrt{3},1+\sqrt{3})$, $k=1,2,3$. Let $
X_{1i}=W_{1i},S_{1i}=\exp(W_{2i}/2),S_{2i}=\log(W_{3i}+1),S_{3i}=\sin(3*(W_{1i}))$. The propensity model is set as $
\operatorname{logit}\{\mathrm{pr}(T_i=1\mid X_i,S_i)\}=0.25X_{1i}-0.25S_{1i}+0.25S_{2i}-0.25S_{3i}$. Finally set $Y_i(1)=0.5X_{1i}-0.5S_{1i}+0.5S_{2i}-0.5S_{3i}+\epsilon_i(1)$,$Y_i(0)=0.5X_{1i}-0.5S_{1i}+0.5S_{2i}-0.5S_{3i}+\epsilon_i(0)$, where $\epsilon_i(1)\sim \mathcal{N}(0,2^2)$, $\epsilon_i(0)\sim \mathcal{N}(0,1)$ and they are independent. 

We estimate the QTE at the $0.5$th and $0.75$th levels. We compare our FQTE $\hat{\Delta}$ using the entire dataset with the initial DR estimators $\hat{\Delta}^{\mathcal{V}}_{p}$ based solely on the validation data. For each fused estimator, we consider three candidate calibrating quantiles, which are $p_{\mathrm{cal},1}=p$, $p_{\mathrm{cal},2}=(0.5,0.75)$ and $p_{\mathrm{cal},3}=(0.25,0.5,0.75)$ for $p=0.5,0.75$ respectively. 

Wald-type 95\% confidence intervals are constructed based on the variance estimates to compare the empirical coverage rates. Sample sizes $(N,n)$ vary from $(2000,500)$,$(2000,1000)$ to $(5000,1000)$ to show effects of increasing $N$ and $n$ respectively. The results in each scenario are based on 2000 replications. We use ``method\_v'' to stand for the initial QTE estimator based on the validation data only, and ``method\_$c_i$'' ($i=1,2,3$) for our fused estimators based on the pseudo quantile $p_{\mathrm{cal},i}$. We also use ``dr11'', ``dr10'', ``dr01'', ``dr00'' to represent the DR estimators with both OR and PS models correctly specified, only the OR model misspecified, only the PS model misspecified, and both misspecified respectively.

\begin{figure}[t!]
	\centerline{\epsfig{file=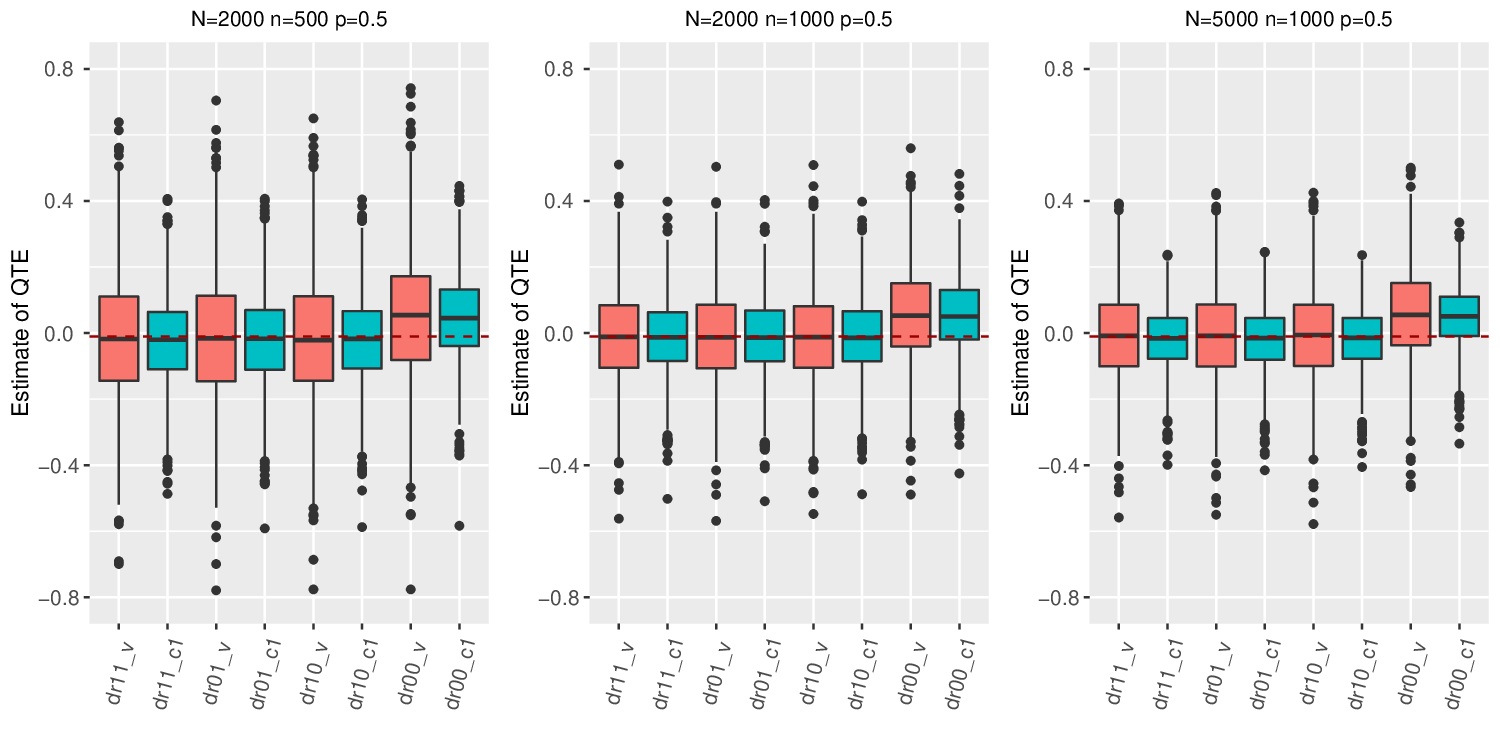,width=4.5in}}\par
	\caption{Point estimates of the DR estimators for $\Delta_{0.5}$. Here ``dr\_v'' represents the initial DR QTE estimator and ``dr\_$c_1$'' represents the DR FQTE using a single quantile.\label{figure1}}
\end{figure}

\begin{table}[t!] 
	\caption{Simulation Results for estimating $\Delta_{0.5}$. Here ``dr\_v'' represents the initial DR QTE estimator and ``dr\_$c_1$'' represents the DR FQTE using a single quantile.}
	\label{table1}\par
	\resizebox{\linewidth}{!}{
			\begin{tabular}{ccccccccccccc}\hline\hline
				$\Delta_{0.5}$      & \multicolumn{4}{c}{$N=2000,  n=500$}  & \multicolumn{4}{c}{$N=2000,  n=1000$}  & \multicolumn{4}{c}{$N=5000,  n=1000$} \\\hline
				Method   & BIAS   & MSE    & SE     & CR     & BIAS    & MSE    & SE     & CR     & BIAS   & MSE    & SE     & CR     \\\hline
				dr11\_v  & 0.0091 & 0.0358 & 0.2017 & 0.9565 & 0.0001  & 0.0185 & 0.1406 & 0.9610 & 0.0028 & 0.0188 & 0.1407 & 0.9535 \\
				\textbf{dr11\_c1} & 0.0131 & 0.0172 & 0.1418 & 0.9655 & 0.0021 & 0.0119 & 0.1139 & 0.9560 & 0.0063 & 0.0083 & 0.0943 & 0.9605 \\\hline
				dr01\_v  & 0.0073 & 0.0373 & 0.2042 & 0.9590 & 0.0005  & 0.0191 & 0.1424 & 0.9580 & 0.0024 & 0.0193 & 0.1424 & 0.9560 \\
				\textbf{dr01\_c1} & 0.0112 & 0.0186 & 0.1453 & 0.9640 & 0.0016 & 0.0126 & 0.1160 & 0.9600 & 0.0067 & 0.0089 & 0.0969 & 0.9560 \\\hline
				dr10\_v  & 0.0080 & 0.0361 & 0.2022 & 0.9605 & 0.0005 & 0.0185 & 0.1410 & 0.9610 & 0.0031 & 0.0189 & 0.1411 & 0.9520 \\
				\textbf{dr10\_c1} & 0.0119 & 0.0173 & 0.1425 & 0.9665 & 0.0027 & 0.0120 & 0.1144 & 0.9570 & 0.0060 & 0.0083 & 0.0948 & 0.9670 \\\hline
				dr00\_v  & 0.0566 & 0.0401 & 0.2045 & 0.9540 & 0.0657  & 0.0235 & 0.1425 & 0.9325 & 0.0678 & 0.0238 & 0.1426 & 0.9300 \\
				\textbf{dr00\_c1} & 0.0537 & 0.0196 & 0.1399 & 0.9500 & 0.0637  & 0.0163 & 0.1137 & 0.9160 & 0.0591 & 0.0114 & 0.0923 & 0.9015 \\\hline\hline
			\end{tabular}}
\end{table}

We find that a one-dimensional calibrating quantile is adequate for efficiency improvement, so here we only display the results with a single calibrating quantile. See the supplementary materials Section S4 for additional simulation details and results.

Figure \ref{figure1} displays the boxplots of 8 FQTEs for ${\Delta}_{0.5}$, with the red boxes representing the initial estimators and green boxes representing the FQTEs. The red dash line in the plots represents the true value. Table \ref{table1} displays the absolute average bias (BIAS), mean squared error (MSE), standard error (SE) calculated by the variance estimates, and the coverage rate (CR) of our Wald-type 95\% confidence intervals for ${\Delta}_{0.5}$. Results for estimating ${\Delta}_{0.75}$ are similar and put in the supplementary materials. Our FQTEs enjoy a large gain of efficiency compared to the initial estimators. Specifically, when $N=2000$ and $n=500$, the MSEs are reduced by half and the SEs are reduced by a third after data fusion. The BIASes increase slightly after data fusion, which may be caused by the estimation of variances. However, since the bias term is ignorable compared to the variance term, we still benefit from data fusion since the SEs are largely decreased. The efficiency gains grow with the sample size of the main data $N$ and become implicit when the sample size $n$ of the validation data is comparable to $N$. This shows that an a larger auxiliary dataset will help more with the efficiency improvement while its role becomes unimportant when the validation dataset is already large enough, which is in line with common sense. 

For ``dr11'', ``dr10'' and ``dr01'', where at least one model is correctly specified, the coverage rates are all around 95\%, which suggests the consistency of our variance estimators. The ``dr00'' estimators are biased since both models are misspecified, and their coverage rates are lower than 95\%.

\section{Application}
\label{sec:application}
In this section, we apply our data fusion method to evaluate the causal effect of smoking during pregnancy on birth weight \citep{Abrevaya2001TheEO,almond2005costs,xie2020multiply}. Based on the Natality Data Set published by National Center for Health Statistics, \citet{almond2005costs} showed that births of low-birthweight babies result in both economic costs for society and the children themselves. Meanwhile, they reported a reduction of 203.2 grams in birthweight for smokers versus nonsmokers. 

Following \citet{Abrevaya2001TheEO} and \citet{almond2005costs}, we focus on the sample of singleton births and mothers who were either white or black, between the ages of 18 and 45, and the residents in Pennsylvania. We limit the sample to infants born in March, June, September, or December. Analysis of other months yields nearly identical results. The resulting sample size is $N=29958$. The treatment variable $T$ here is the mother's smoking status during pregnancy, and the outcome variable $Y$ here is the birthweight of infants (in grams). There are 5558 smokers and 24400 nonsmokers in total. Due to economic costs, researchers may be more interested in the causal effect in the lower quantiles of birthweight. So we consider estimating the $0.5$th and $0.25$th QTE of smoking during pregnancy on birth weight. 

At the same time, large surveys cost a lot of money and time to follow up with the participants and collect some important measures. It may be of great value to cut down the sample size of data needed with full confounders. To illustrate the validity of our data fusion method, we construct the main dataset by including only the basic confounders: mother's marital status, mother's race (either black or white), gender of the infant, mother's age, mother's education and the number of prenatal visits. These five confounders are used as full confounders to evaluate the $0.5$th QTE in \citet{xie2020multiply}. However, there are additional key confounders not included in their analyses: alcohol use during pregnancy, the average number of drinks per week, and adequacy of care. We construct the validation dataset by selecting random samples from the whole data including all these eight confounders. The sample size $n$ of the validation dataset varies from $2000,5000$ to $10000$. With estimates based on the whole data with all confounders as a benchmark, we compare our fused estimators using both the validation and main datasets with the initial estimators based solely on the validation dataset. For each fused estimator, we take $p_{\operatorname{cal}}$ the same as $p$ for calibration and we propose a normal linear model for the outcome and a logistic model for the propensity score. We use the random forest to estimate the propensity score in density estimation. The results for QTE estimation are displayed in Table \ref{table3}.

\begin{table}[t!] 
	\caption{Point Estimate (Est), Standard Error (SE) and the Wald-type 95\% Confidence Interval. Here ``dr\_v'' represents the initial DR QTE estimator and ``dr\_$c_1$'' represents the DR FQTE using a single quantile.}
	\label{table3}\par
	\resizebox{\linewidth}{!}{\begin{tabular}{cccccccc}\hline\hline
			\multicolumn{2}{c}{}       & \multicolumn{3}{c}{$\Delta_{0.5}$}                 & \multicolumn{3}{c}{$\Delta_{0.25}$}                  \\\hline
			$n$&  Method   & Est     & SE    & 95\%CI      & Est     & SE    & 95\%CI         \\\hline
			$2000$  & dr\_v  & -226    & 44.25   & {[}-312.73,-139.27{]}    & -227    & 43.66 & {[}-312.58,-141.42{]} \\
			& {dr\_c1} & -206.95 & 21.75   & {[}-249.59,-164.33{]}    & -219.84 & 21.16 & {[}-261.32,-178.39{]} \\\hline
			$5000$  & dr\_v  & -189    & 25.68 & {[}-239.32,-138.68{]}    & -182    & 28.92 & {[}-238.68,-125.32{]} \\
			& {dr\_c1} & -189.59 & 15.15   & {[}-219.22,-159.82{]} & -215.30 & 19.45 & {[}-253.81,-177.57{]} \\\hline
			$10000$ & dr\_v  & -200    & 17.90   & {[}-235.09,-164.91{]}    & -199    & 20.57 & {[}-239.33,-158.67{]} \\
			& {dr\_c1} & -193.24 & 11.08   & {[}-214.93,-171.48{]}    & -205.98 & 13.19 & {[}-231.89,-180.19{]} \\\hline
			$N$     & dr\_v  & -198    & 10.05   & {[}-217.70,-178.30{]}    & -205    & 11.63 & {[}-227.80,-182.20{]} \\\hline
	\end{tabular}}
\end{table}

As we can see from Table \ref{table3}, the DR estimator for the $0.5$th QTE based on the whole sample using all the eight confounders is $-198$, which indicates a reduction of $-198$ grams in the median of birthweight for smokers and nonsmokers. Similarly, a larger reduction in the 0.25th quantile of the birthweight, which is $-205$ grams, is reported. What we want to emphasize here is that our FQTE greatly improves the efficiency of the initial estimators. Therefore, the length of confidence intervals after data fusion is reduced significantly. When the validation sample size $n$ is $10000$, which is approximately a third of the whole sample size $N$, the standard error after data fusion is nearly comparable to that of the estimators based on the whole sample with full confounders. This indicates that we can only collect important information on a representative subsample of the whole data and then our data fusion method can still provide as efficient QTE estimator as that obtained by the whole fully-observed data. Consequently, the cost of money and time in large surveys can be greatly reduced.

\section{Discussion}
\label{sec:discussion}
\subsection{Generalization of the projection idea}
What lies at the heart of our method is the connection of two datasets through estimating functions, which is also a consistent estimator for zero. Once these estimating functions are obtained, projection can be done to improve the efficiency of initial estimators. Inspired by this, consider $\gamma^{\operatorname{Conf}}$ to be an $d$-dimensional vector of parameters identifiable on the joint distribution of $U=(Y,T,X)$ with
\begin{equation}\label{generalconf}
E\left\{\varphi(U;\gamma^{\operatorname{Conf}},\zeta^{\operatorname{Conf},*})\right\}=0,
\end{equation}
where $\zeta^{\operatorname{Conf},*}$ is an unknown nuisance parameter. Note that (\ref{Confemdr}) is a special case of (\ref{generalconf}) by taking $\varphi=\phi_t$, $\zeta^{\operatorname{Conf},*}$ as $q_{t,p}^{\operatorname{Conf}}$ and $\zeta^{\operatorname{Conf},*}$ as $\eta_t^{\operatorname{Conf},*}$. Under Assumption \ref{assumption5}, based on the estimates $\hat\zeta^{\operatorname{Conf}}$ and $\hat\gamma^{\operatorname{Conf}}$ obtained from the main data, we can then connect the two datasets based on estimating functions similar to that in (\ref{Ct}), formed as
$\hat{C}={1}/{n}\sum_{i=1}^{n}\varphi(U_i;\hat\gamma^{\operatorname{Conf}},\hat\zeta^{\operatorname{Conf}})$.
As implied by Theorem \ref{theorem1}, the intuition is to choose $\varphi$ which results in a $\hat{C}$ with a larger variance and as correlated with the initial estimator as possible. The choice of a well-designed $\varphi$ to make larger efficiency improvement may be discussed in further work. 

\subsection{Relaxation of Assumption \ref{assumption3}}\label{subsec:assump3}
We may weaken Assumption \ref{assumption3} to the correct specification of the conditional quantile with a small modification of the DR estimation equation (\ref{emdr}). Consider the quantile regression model $Y_i=g_t(X,S;\theta_t)+\epsilon_{t,i}, $ for $T_i=t$ $(t=0,1)$, where $g_t$ is known with parameter $\theta_t\in\mathbb{R}^{p_x+p_s}$ and $P(\epsilon_{t,i}<0\mid X_i,S_i)=p$. Given an estimate $\hat{\theta}_t$ and the residuals $\hat{\epsilon}_{t,i}=Y_i-g_t(X_i,S_i;\hat\theta_t)$, we can replace Equation (\ref{emdr}) with 
\begin{equation*}
{1}/{n} \sum_{i=1}^{n} \Psi^{\mathrm{mod}}_t(O_i;q,\hat{\eta}_{t})= 0,
\end{equation*}
where $\Psi^{\mathrm{mod}}_t(O_i;q,\hat{\eta}_{t})=\hat w_{t,i}\{I{(Y_{i} \leqslant q)}-\frac{1}{n_t}\sum_{i:T_i=t}I(\hat{\epsilon}_{t,i}\leq q)\}+\frac{1}{n_t}\sum_{i:T_i=t}I(\hat{\epsilon}_{t,i}\leq q)-p$ and $n_t=\sum_{i=1}^{n}I{(T_i=t)}$ . The modified quantile estimator is also doubly robust in the sense that it is consistent if either the propensity score model or the conditional quantile $g_t(X,S;\theta_t)$ is correctly specified. Asymptotic linear representations can be also established in a parallel way with a more rigorous proof. Similar ideas can be found in \cite{sued2020robust}.

\subsection{Multiple Auxiliary Datasets}
We may also extend our method to incorporate multiple auxiliary datasets as that in \citet{yang2020combining}. Specifically, let $U^{(k)}=(Y,T,M^{(k)\top})$ denote the partially-observed information on the $k$-th auxiliary dataset for $k=1,\ldots,K$, where $M^{(k)}\subsetneqq (X^{\top},S^{\top})^{\top}$. Let $\mathcal{O}^{(k)}$ denote the index set of the $k$-th auxiliary dataset with sample size $N_k$. For the $k$-th dataset, we inherit notations from Section 2.3 with $X$ replaced by $M^{(k)}$ and add superscript $(k)$. Following similar procedure as that in Section \ref{subsec:partial}, we can obtain confounded DR quantile estimators $\hat{q}^{\mathrm{Conf},(k)}_{t,p}$ by solving the corresponding estimation equation $1/N_k\sum_{i\in\mathcal{O}^{(k)}}\phi^{(k)}_t(U_i^{(k)};q,\hat{\eta}^{\mathrm{Conf},(k)}_{t})=0$. Then similar to (\ref{Ct}), we can integrate the information from the $k$-th auxiliary dataset through $\hat{C}_t^{(k)}=1/n\sum_{i=1}^{n}\phi^{(k)}_t(U_i^{(k)};q^{\mathrm{Conf},(k)}_{t,p},\hat{\eta}^{\mathrm{Conf},(k)}_{t})$ for $t=0,1$. Asymptotical linear representations for $\{\hat{C}_{t}^{(k)}\}$ can be established based on $\phi^{(k)}_{t,i}$. Let $\hat{C}^{(k)}=(\hat{C}^{(k)\top}_{1},\hat{C}^{(k)\top}_{0})$ and $\hat{C}=(\hat{C}^{(1)\top},\ldots,\hat{C}^{(K)\top})$. We can also obtain (\ref{n1qte}) with $\varrho$ and $\Sigma_{\mathrm{ep}}$ depending on $\psi_{t,i}$ and $\{\phi_{t,i}^{(k)}\}_{k=1}^{K}$. Then the FQTE integrating multiple datasets can be similarly obtained by (\ref{qte}).

\subsection{Sensitivity Analysis}
In the real world, the heterogeneity may be intrinsic between the two samples, and hence $\hat{C}_t$ in (\ref{Ct}) may not converge to 0. Inspired by \citet{yang2020combining}, we can introduce a sensitivity parameter $\delta$ to quantify the systematic heterogeneity and replace (\ref{n1qte}) with 
\begin{equation*}
n^{1 / 2}\left(\begin{array}{c}
\hat{\Delta}^{\mathcal{V}}_{p}-\Delta_{p} \\  \hat{C}-\delta
\end{array}\right) {\longrightarrow} \mathcal{N}\left\{0,\left(\begin{array}{cccc}
\sigma_{\mathcal{V}}^{2} & {\varrho}^{\top} \\
\varrho&\Sigma_{\mathrm{ep}}
\end{array}\right)\right\}.
\end{equation*}
The modified estimator becomes $\hat{\Delta}^{\mathrm{mod}}_{p}(\delta)=\hat{\Delta}^{\mathcal{V}}_{p}-\hat{\varrho}^{\top} \hat{\Sigma}_{\mathrm{ep}}^{-1}(\hat{C}-\delta)$. In this way, an investigator is able to assess the impact of the heterogeneity of the two data by varying the values of $\delta$.

\section*{Supplementary Materials}
The supplementary materials contain extensions to allow for a missing at random mechanism, technical proofs, and additional simulation results.
\par
\section*{Acknowledgements}
We sincerely thank the editor, associate editor, and two anonymous reviewers for their insightful comments. This work is partially supported by the National Natural Science Foundation of China grants 12071087.
\par

\bibhang=1.7pc
\bibsep=2pt
\fontsize{9}{14pt plus.8pt minus .6pt}\selectfont
\renewcommand\bibname{\large \bf References}
\expandafter\ifx\csname
natexlab\endcsname\relax\def\natexlab#1{#1}\fi
\expandafter\ifx\csname url\endcsname\relax
\def\url#1{\texttt{#1}}\fi
\expandafter\ifx\csname urlprefix\endcsname\relax\def\urlprefix{URL}\fi

\bibliographystyle{Chicago}
\bibliography{ref}

\vskip .65cm
\noindent
Department of Statistics and Data Science, Fudan University
\vskip 2pt
\noindent
E-mail: (yijiaozhang20@fudan.edu.cn)
\vskip 2pt

\noindent
Department of Statistics and Data Science, Fudan University
\vskip 2pt
\noindent
E-mail: (zhuzy@fudan.edu.cn)

\end{document}